\documentclass[aps,prb,superscriptaddress,twocolumn]{revtex4-1}

\usepackage{todonotes}

\usepackage{times}
\usepackage{amsmath}
\usepackage{amsfonts}
\usepackage{amssymb}
\usepackage{epsfig}
\usepackage{siunitx}

\newcommand{\md}{\mathrm{d}}
\newcommand{\me}{\mathrm{e}}
\newcommand{\mi}{\mathrm{i}}

\begin{document}

\title{Extracting single electron wavefunctions from a quantum electrical current }

\author{A. Marguerite$^{1}$, B. Roussel$^{2}$, R.
Bisognin$^{1}$, C. Cabart$^{2}$, M. Kumar$^{1}$, J.-M. Berroir$^{1 }$,  E. Bocquillon$^{1 }$, B.
Pla\c{c}ais$^{1 }$,   A. Cavanna$^{3}$, U.
Gennser$^{3 }$, Y. Jin$^{3 }$,  P. Degiovanni$^{2}$, and G. F\`{e}ve$^{1 \ast}$ \\
\normalsize{$^{1}$Laboratoire Pierre Aigrain, Ecole Normale
Sup\'erieure-PSL Research University, CNRS,}
\normalsize{Universit\'e Pierre et Marie
Curie-Sorbonne Universit\'es, Universit\'e Paris Diderot-Sorbonne Paris
Cit\'e,}\\
 \normalsize{24 rue Lhomond, 75231 Paris Cedex 05, France.}\\
\normalsize{$^{2}$ Univ Lyon, Ens de Lyon, Universit\'e Claude Bernard
Lyon 1, CNRS,}\\
\normalsize{Laboratoire de Physique, F-69342 Lyon, France.}\\
\normalsize{$^{3}$ Centre de Nanosciences et de Nanotechnologies, CNRS, Univ. Paris-Sud, Universit\'{e} Paris-Saclay,}\\ \normalsize{C2N-Marcoussis,
91460 Marcoussis, France.}
\\ \normalsize{$^\ast$ To whom correspondence should
be addressed; E-mail:  feve@lpa.ens.fr.}\\
}

\begin{abstract}
Quantum nanoelectronics has entered an era where quantum electrical
currents are built from single to few on-demand elementary excitations.
To date however, very limited tools have been implemented to
characterize them. In this work, we present a quantum current analyzer
able to extract single particle excitations present within a periodic
quantum electrical current without any a priori hypothesis.
Our analyzer combines two-particle interferometry and signal processing
to extract the relevant electron and hole wavefunctions localized around
each emission period and their quantum coherence from one
emission period to the other. This quantum current analyzer opens new
possibilities
for the characterization and control of quantum electrical currents in nanoscale conductors and
for investigations of entanglement in quantum electronics down to the
single electron level.
\end{abstract}

\maketitle

Can we extract the full single particle content of a quantum electrical current? Although growing in importance by the availability of on-demand single to few electron
sources\cite{Ondemand,Leicht2011,Dubois2013,
Fletcher2013}, characterizing the excitations of a quantum electrical
current is still an open problem.
In the stationary regime,
electronic transport is described in term of coherent scattering of
electronic modes characterized by their energies\cite{Buttiker85}.
The knowledge of the mode population given by the
electronic distribution function is
then sufficient to predict any single-particle physical quantity.
In the non-stationary regime, slow drives generate quasi-classical
currents carrying a large number
of excitations per period, which can also be viewed as adiabatic evolutions
of the d.c. case. Conversely, fast drives lead to fully quantum electrical currents that can
be used to transmit information carried by single to few electronic excitations
transferred coherently during each period of the source.
From a basic quantum
physics perspective, as well as in terms of quantum technological applications, it is then natural to ask
what their wavefunctions and what their injection
probabilities are.

Beyond time\cite{Ondemand,Kataoka2016} and energy
distribution measurements\cite{Altimiras2009,Fletcher2013}, which are not able to answer these
questions completely, theoretical works have investigated the elementary events of charge transfer through a quantum point contact\cite{Vanevic2007}. Recent experimental works have
probed electron and hole wavefunctions in a tunnel junction under a.c. excitation\cite{Vanevic2016} or have reconstructed the
electronic Wigner distribution\cite{Jullien2014} of a Leviton excitation\cite{Levitov96,Keeling06,Dubois2013}. However, in both
cases, the characterization of the input state is based on strong
assumptions and cannot extract the electron and hole wavefunctions from an arbitrary
quantum electrical current.

In this article, we present an on-chip quantum electrical current analyzer combining
two-particle interferometry \cite{Liu1998,Neder2007,Ol'khovskaya2008,Bocquillon2013} with signal processing
techniques to answer these questions. The single particle content of a
quantum electrical current is encoded within the electronic Wigner
distribution\cite{DegioWigner2013,Kashcheyevs2017} which is a very relevant tool for describing
quantum states, in particular of light\cite{Bertet2002}, atoms\cite{Leibfried1996} or molecules\cite{Dunn1995}.
Using sinusoidal drives for
demonstration, we reconstruct the Wigner distributions of both a
quasi-classical and a quantum electrical current. In the latter case, we extract the
single particle wavefunctions from the reconstructed Wigner distribution
as well as their coherence across different periods. This experiment demonstrates
our ability to extract the full single particle content of a quantum
current from experimental data without any assumption on
the electronic state, opening new perspectives for electron and microwave quantum optics in conjunction with quantum information in quantum conductors.

\begin{figure}[h!] %
\includegraphics[width=\linewidth]{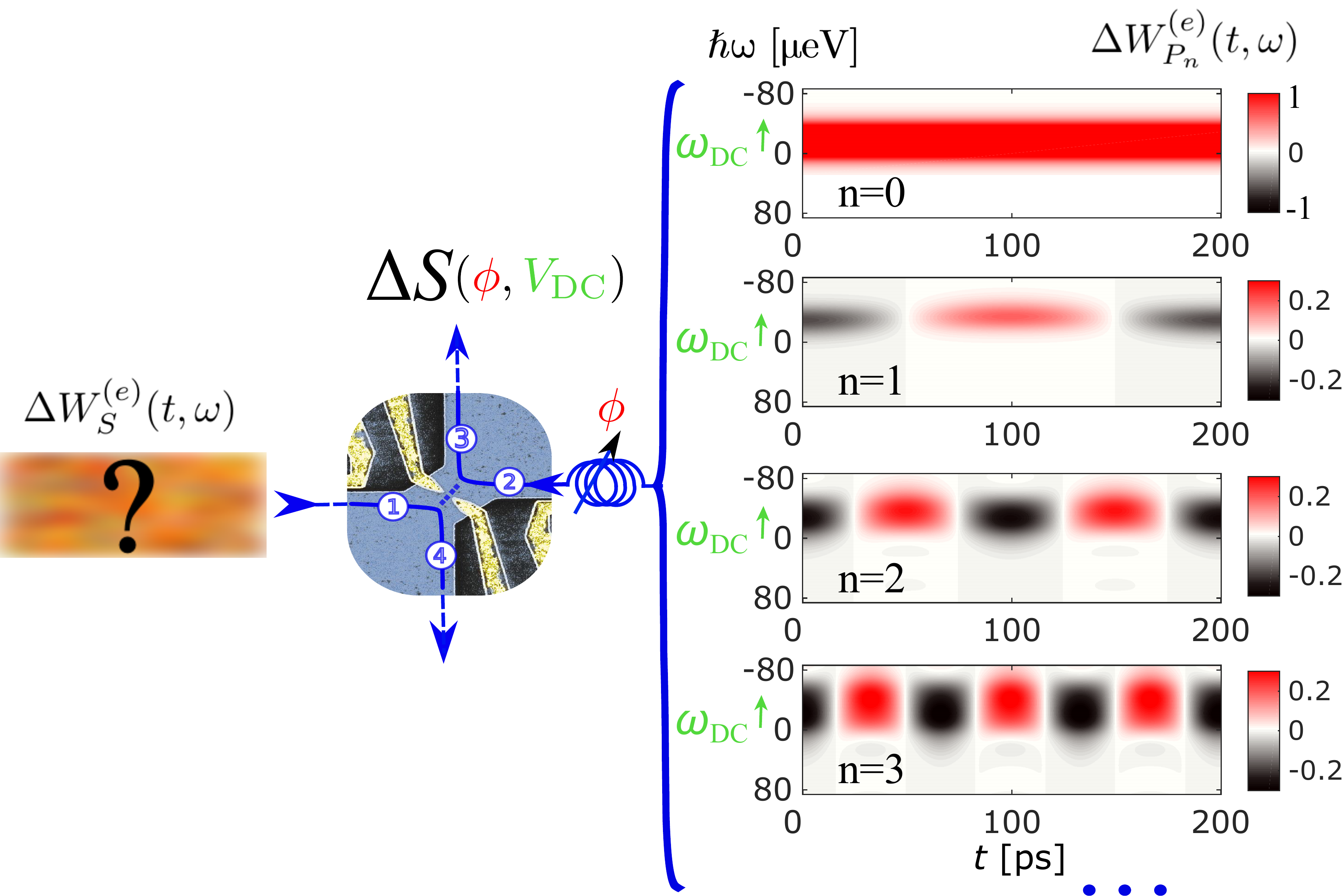}
\caption{  The unknown
Wigner distribution $\Delta W^{(e)}_S(t,\omega)$ is sent into input 1. The
probe signals $\Delta W^{(e)}_{P_n}(\omega)$ sent into input
2 are plotted for $n=0$ to $n=3$, time (energy) is on the horizontal (vertical) axis. The frequency is $f=\SI{5}{\giga\hertz}$ and the
temperature $T_{\text{el}}=\SI{80}{\milli\kelvin}$). The excess noise
$\Delta S$ is measured  in output 3 as a function of the d.c. bias $V_{\text{DC}}=-\hbar\omega_{\text{DC}}/e$ and the phase
difference $\phi$ between the source and probe signals.}
\label{fig2}
\end{figure}

Similarly to the electronic distribution function $f(\omega)$, which describes the occupation probability of electronic modes as a function of energy $\hbar \omega$ and contains all the single-body properties of a stationary current, the Wigner distribution $W^{(e)}(t,\omega)$ encodes all the single-particle properties of the electronic state in the non-stationary case. It also provides a direct way to distinguish between quasi-classical and quantum currents\cite{DegioWigner2013}. Quasi-classical currents are characterized by bounded values of the Wigner distribution, $0 \leq W^{(e)}(t,\omega)
\leq 1$, such that $W^{(e)}(t,\omega)$ can be interpreted as a time-dependent electronic distribution function. Conversely, deviations from these classical bounds are the hallmark of the quantum regime where $ W^{(e)}(t,\omega)$ can be used to extract electron and hole wavefunctions.
Following the protocol presented in Ref.\cite{Tomo2011}, an unknown
Wigner distribution can be reconstructed using an electronic Hong-Ou-Mandel\cite{HOM}
interferometer \cite{Ol'khovskaya2008,Bocquillon2013}, which measures the overlap between two
electronic states propagating towards the two inputs of a beam-splitter. The interferometer is implemented in a
two-dimensional electron gas in the quantum Hall regime at filling
factor $\nu=2$. A quantum point contact is used as an electronic
beam-splitter partitioning the outer edge channel. As can be
seen on the sketch of the experiment represented in Figure \ref{fig2},
the source generating the unknown state labeled by the subscript $S$ is propagating in input 1
whereas input 2 is fed with a set of reference states called
probe states and labeled $\{P_n\}, n \in \mathbb{N}$.
The low frequency current noise at the output of the splitter probes the degree of indistinguishability between the source and probe states. To isolate the contribution of the source, we measure the
excess noise $\Delta S$ at splitter output 3 between the on and off states of the source:
\begin{align}
 \Delta S &= 2 e^2 \mathcal{T}(1-\mathcal{T}) \int
 \frac{\md\omega}{2\pi}\left[ \overline{\Delta W^{(e)}_S}^t
 \left(1-2f_{\text{eq}}\right)
 \right.  \nonumber \\
 & \qquad \qquad \qquad \qquad \qquad \qquad -\left.
 2\overline{\Delta W^{(e)}_S\Delta W^{(e)}_{P_n}}^t \right]
 \label{SHOM}
\end{align}
where $\mathcal{T}$ is the beam-splitter transmission,
$\overline{\cdots}^t$ denotes the average over time $t$ and $\Delta
W^{(e)}_{S/P_{n}}$
are respectively the source and probe excess Wigner distribution with
respect to the equilibrium situation described by the Fermi-Dirac
distribution: $W^{(e)}_{S/P_{n}}(t,\omega)=f_{\text{eq}}(\omega) +
\Delta W^{(e)}_{S/P_{n}}(t,\omega)$. The first term in Eq.\eqref{SHOM} represents
the classical random partition noise of the source. It is reduced by
the second term which represents two-particle interferences between
source and probe related to the overlap between $\Delta W^{(e)}_S$ and
$\Delta W^{(e)}_{P_{n}}$. By properly choosing the probe states, Eq.\eqref{SHOM}
allows for the reconstruction of any unknown Wigner distribution
\cite{Tomo2011,DegioWigner2013}.

\begin{figure}[h] %
\includegraphics[width=\linewidth]{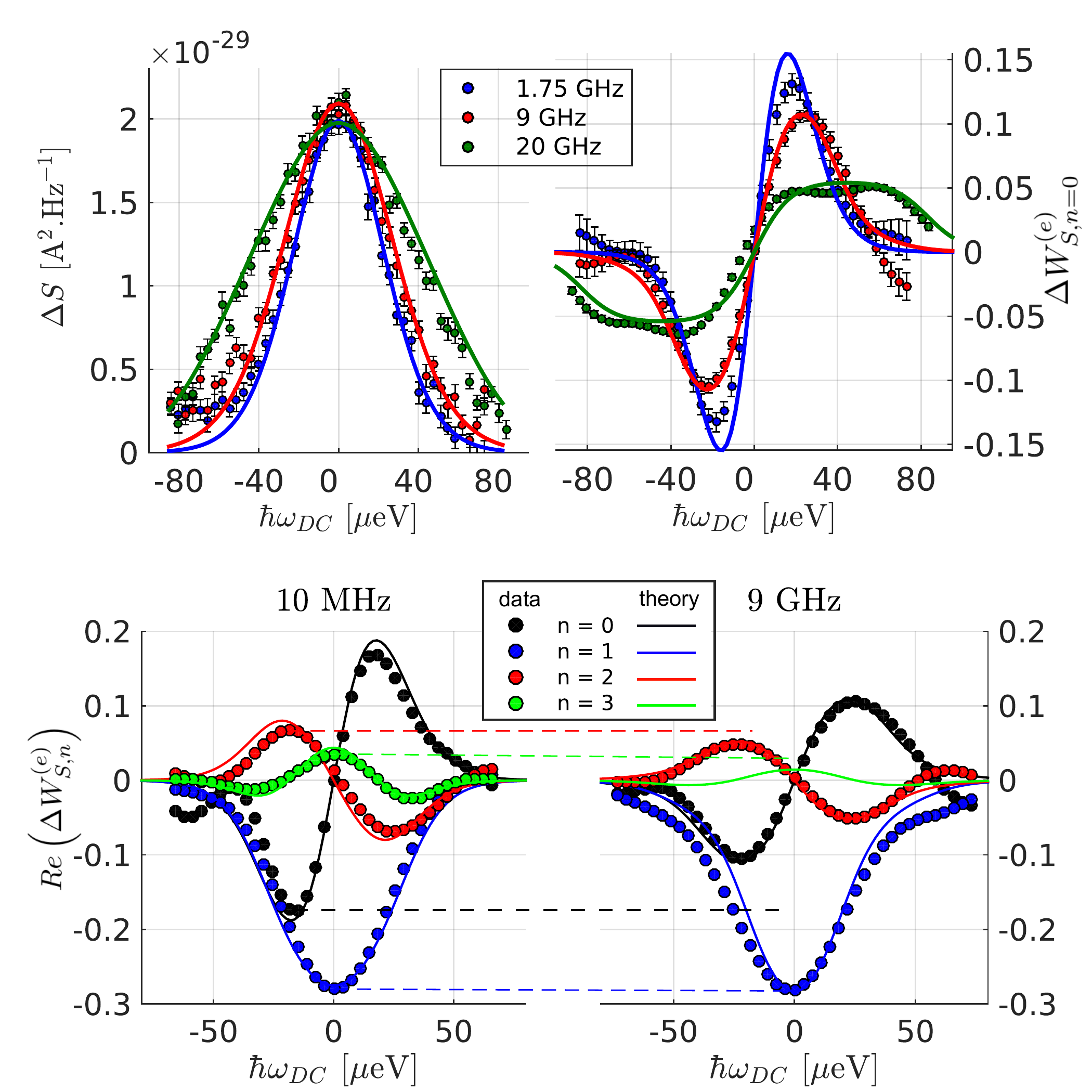}
\caption{ Upper
left panel: Excess noise $\Delta S$ as a function of d.c bias
$\omega_{\text{DC}}$. The plain lines represent numerical calculations
for $V_{S}=\SI{28}{\micro\volt}$, $f=\SI{1.75}{\giga\hertz}$,
$V_{S}=\SI{31}{\micro\volt}$ , $f=\SI{9}{\giga\hertz}$, and
$V_{S}=\SI{38}{\micro\volt}$, $f=\SI{20}{\giga\hertz}$ and
$T_{\text{el}}=\SI{100}{\milli\kelvin}$. Upper right panel:
electronic distribution function $\Delta W^{(e)}_{S,0}(\omega)$. Plain lines
represent numerical calculations (same parameter as the left
panel). Lower panel: $\Delta W^{(e)}_{S,n}(\omega)$
for $f=\SI{10}{\mega\hertz}$ (left) and $f=\SI{9}{\giga\hertz}$ (right).
The plain lines represent numerical calculations with
$T_{\text{el}}=\SI{100}{\milli\kelvin}$, $V_S=\SI{33}{\micro\volt}$
($f=\SI{10}{\mega\hertz}$) and $V_S=\SI{31}{\micro\volt}$ ($f=\SI{9}{\giga\hertz}$).}
\label{fig3}
\end{figure}

As a proof of concept of our quantum current analyzer, we use simple
sinusoidal drives, $V_S(t)=V_{S}\cos{(2 \pi f t)}$ at various
frequencies $f=1/T$, a convenient
choice which does not affect the generality of our results. In this
case, the low frequency drives with $hf \lesssim k_{B}T_{\text{el}}$
(where $T_{\text{el}}=\SI{100}{\milli\kelvin}$ is the electronic temperature) are in the quasi-classical
regime. $W^{(e)}_S$ should then be given by a Fermi sea with a time
dependent chemical potential following the a.c. drive:
$W^{(e)}_S(t,\omega) = f_{\text{eq}}\left(\omega+\frac{eV_{S}}{\hbar}\cos{(2 \pi
f t)}\right)$. Conversely, $hf \gtrsim k_{B}T_{\text{el}}$ corresponds
to the quantum regime where deviations from the classical bounds are expected.
As the source state is periodically
emitted, a convenient choice of
probe states basis is suggested by a Fourier expansion of the source Wigner
distribution: $\Delta W^{(e)}_S(t,\omega) = \sum_{n} \Delta W^{(e)}_{S,n}(\omega)
\me^{2 \pi \mi n f t}$.
Let us first focus on the $n=0$ harmonic, $\Delta W^{(e)}_{S,0}(\omega)$, which
represents the source excess electronic distribution function. This time averaged
quantity can be extracted from
Eq.\eqref{SHOM}  by using a d.c. bias $V_{\text{DC}}$ as the probe $P_0$
(see Fig.\ref{fig2}) so that
$\Delta W^{(e)}_{P_0}(t,\omega)=f_{\text{eq}}\left(\omega
-\omega_{\text{DC}}\right) - f_{\text{eq}}(\omega)$ (with
$\omega_{\text{DC}}=-eV_{\text{DC}}/\hbar$) is very close to $1$ for
$0\leq \omega \leq \omega_{\text{DC}}$ and to $0$ elsewhere.
Fig.\ref{fig3} (upper left panel) represents the excess noise
$\Delta S$ as a function of $\omega_{\text{DC}}$ for various
sinusoidal source drives of increasing
frequency (\SIlist{1.75;9;20}{\giga\hertz}) with drive
amplitudes ($V_{\text{s}}=$ \SIlist{28;31;38}{\micro\volt} respectively) chosen to produce similar
partition noise (around $\SI{2e-29}{\ampere\squared\per\hertz}$)
at $\omega_{\text{DC}} =0$ ($P_0$ switched off).
The partition noise is suppressed to zero by two-electron interferences when the d.c. bias is
increased but in different ways for different frequencies.
$\Delta
W^{(e)}_{S,0}(\omega)$ can then be obtained via the derivative of $\Delta S$ with
respect to $\omega_{\text{DC}}$\cite{Tomo2011, Gabelli2013} (see Supplementary Material). The obtained values of $\Delta
W^{(e)}_{S,0}(\omega)$ are plotted on the upper right panel of Figure \ref{fig3}.
The three curves show that as the drive
frequency increases, the spectral weight is shifted towards higher
frequencies. In the quasi-classical case ($f=\SI{1.75}{\giga\hertz}$),
the typical energy at which electrons (respectively holes)
are promoted above (respectively below) the Fermi energy is given
by  the amplitude $eV_{S}=\SI{28}{\micro\eV}$ of the chemical potential variations.
On the contrary, in the quantum regime ($f=\SI{20}{\giga\hertz}$ so that
$hf > k_{B}T_{\text{el}}$), electron/hole pair creation results from the
absorption of photons at energy $hf$\cite{Reydellet2003,Rychkov2005}.
The excess electron distribution function presents
square steps of width $h f=\SI{83}{\micro\eV}$ and
amplitude $(eV_{S}/hf)^2/4$, which is the probability to absorb
one photon from the drive. Our experimental results
compare very well with photo-assisted noise calculations
\cite{Rychkov2005} (plain lines) without any adjustable parameters, thus
confirming the robustness of our measurement of $\Delta
W^{(e)}_{S,0}(\omega)$.

Accessing the time dependence of $W^{(e)}_S(t,\omega)$ requires measuring
the non-zero harmonics $\Delta W^{(e)}_{S,n\neq 0}$.
Fortunately,
an accurate reconstruction is usually achieved already with
$|n|\leq 3$. Following Eq.\eqref{SHOM},
accessing the $n$th harmonic requires a probe $P_n$
whose Wigner
distribution $W^{(e)}_{P_n}$ evolves periodically in time at frequency $n f$. At
low excitation amplitude, the $W^{(e)}_{P_n}$ depends linearly on the probe voltage, $V_{P_n}(t)=V_{P_n}\cos{(2 \pi n f t + \phi)}$, chosen to extract
$\Delta W^{(e)}_{S,n}$\cite{DegioWigner2013}:
\begin{align}
\Delta
W^{(e)}_{P_n}(t,\omega) &= -\frac{eV_{P_n}}{\hbar} \cos{\left(2 \pi n f t +
\phi\right)} \; g_n\left(\omega -\omega_{\text{DC}}\right), \label{Wigprobe}
\end{align}
with $g_n(\omega)= \Big(f_{\text{eq}}(\omega-n\pi f) - f_{\text{eq}}(\omega+n\pi f)\Big)/(2\pi nf)$.
$\Delta W^{(e)}_{P_n}$ for $n=1$ to $n=3$ are plotted on Fig.
\ref{fig2}. Changing the phase $\phi$ allows us to scan the temporal axis,
whereas the width of $\Delta W^{(e)}_{P_n}$ along the energy axis is fixed by
the width of $g_n$, given by the maximum of $k_B T_{\text{el}}$ and $nhf$. As in the
$n=0$ case, varying a d.c. bias $V_{\text{DC}}$ on top of the a.c. probe
excitation enables scanning the energy axis.
Using Eqs.~\eqref{SHOM} and \eqref{Wigprobe}, the
real and imaginary parts of $\Delta W^{(e)}_{S,n}$ can be directly related to
the variations of $\Delta S$ as a function of the phase
$\phi$ and d.c. bias $V_{\text{DC}}$ (see Supplementary Material).

\begin{figure}[h!]
\includegraphics[width=\linewidth]{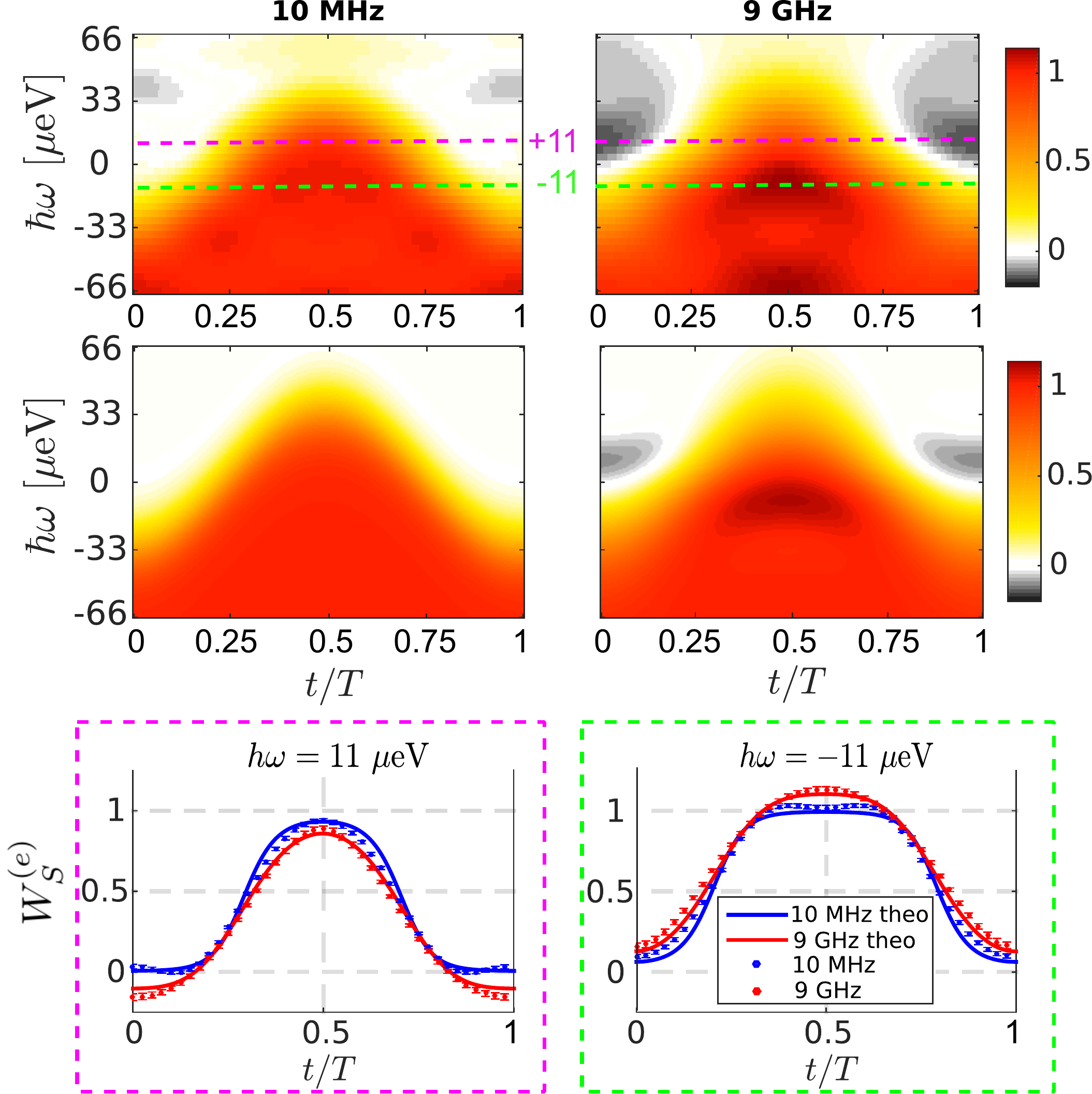}
\caption{Upper
panels: experimental data of $W^{(e)}_S(t,\omega)$ in the quasi-classical
($f=\SI{10}{\mega\hertz}$, left) and quantum ($f=\SI{9}{\giga\hertz}$, right) cases.
Middle panels: theoretical calculations of $W^{(e)}_S(t,\omega)$ at
$f=\SI{10}{\mega\hertz}$ (left) and $f=\SI{9}{\giga\hertz}$ (right). Lower panels: cuts of $W^{(e)}_S(t,\omega)$ at
constant energy $\hbar \omega=\SI{11}{\micro\volt}$ (left) and $\hbar \omega=
-\SI{11}{\micro\eV}$ (right).
Blue points are for $f=\SI{10}{\mega\hertz}$, red points for
$f=\SI{9}{\giga\hertz}$, the plain lines
represent numerical calculations.}
\label{fig4}
\end{figure}

Figure \ref{fig3} presents $\Re{(\Delta W^{(e)}_{S,n})}$ ($\Im{(\Delta W^{(e)}_{S,n})}=0$) for
$n=0,1,2,3$ for a quasi-classical drive $f=\SI{10}{\mega\hertz}$ (lower left panel), and
$n=0,1,2$ ($n=3$ falls below our experimental resolution) for a
quantum drive $f=\SI{9}{\giga\hertz}$ (lower right panel). While the $n=1$ harmonics take
very close values as explained by the similar amplitudes
of the drives ($V_{S}=\SI{33}{\micro\volt}$ for $f=\SI{10}{\mega\hertz}$ and
$V_{S}=\SI{31}{\micro\volt}$ for $f=\SI{9}{\giga\hertz}$),
the $n=0$, $2$ and $3$ harmonics are lower in the high frequency case compared to
the low frequency one. Indeed, these terms are related to multiphoton
absorption/emission processes, whose strength increases with the ratio
$\alpha=eV_{S}/h f$. This ratio is very high in the
quasi-classical case ($\alpha \approx 800$) and smaller than one in the
quantum one ($\alpha \approx 0.8$) thus explaining the smaller amplitude
of the harmonics $n\neq 1$. After extracting all relevant $\Delta
W^{(e)}_{S,n}$, we can combine them to reconstruct the full Wigner
distribution:
\begin{align}
 W^{(e)}_S(t,\omega) &= f_{\text{eq}}(\omega) + \Delta
 W^{(e)}_{S,0}(\omega) \nonumber \\
 & +2\sum_{n=1}^{N}\Re{(\Delta
 W^{(e)}_{S,n}(\omega))}\cos{\left(2
 \pi n f t\right)},
\end{align}
where the sum extends to $N=3$ at $f=\SI{10}{\mega\hertz}$ and $N=2$ at
$f=\SI{9}{\giga\hertz}$, and where
$\Im{(\Delta W^{(e)}_{S,n})} = 0$ has been used.
The two Wigner distributions are represented on
Fig.
\ref{fig4} (upper panel).
Apart for small discrepancies related to the
deconvolution process, the quasi-classical case is very close to the
expected equilibrium distribution function with a time varying chemical
potential $\mu(t)=-eV_{S}\cos{(2\pi ft)}$ (see middle left
panel). In particular, it is basically constrained to values between $0$ and $1$.
In contrast, in the quantum case, the Wigner distribution can
take values that are strongly negative or that are well above one.
Consequently, single-particle properties
are no longer described in terms of a time varying electronic
distribution function, in good agreement with theoretical predictions
(middle right panel). These differences with the classical case can be understood by plotting
cuts of the Wigner distribution at constant energy $\hbar\omega = \pm
\SI{11}{\micro\eV}$
(lower panels of Fig. \ref{fig4}). In the classical case the sizeable values of the
high harmonics of the Wigner distribution are necessary to
reconstruct an equilibrium Fermi distribution which varies sharply from
0 to 1. On the contrary, high harmonics are suppressed in the quantum case such that $W^{(e)}_S(t,\omega)$
varies in a much smoother way. This
explains the overshoots (undershoots)
above $1$ (below $0$) which agree well with theoretical expectations (plain lines).

The second step of our quantum signal dissection scheme extracts individual
electronic wavepackets from the reconstructed Wigner distribution.
This reconstruction is performed on the excess single electron coherence \cite{Bocquillon2014,Haack2013}
$\Delta_0 \mathcal{G}^{(e)}_S(t+\frac{\tau}{2},t-\frac{\tau}{2})$, which is the inverse
Fourier transform with respect to $\omega$ of the excess Wigner distribution $\Delta_0
W^{(e)}_S(t,\omega)=W^{(e)}(t,\omega)-\Theta(-\omega)$,
defined with respect to the zero temperature Fermi electronic distribution function $\Theta(-\omega)$.
This reference choice
ensures that all excitations, including the thermal ones, are extracted by
our algorithm.
The goal is to find the
simplest expression of the excess single electron coherence in terms of
recently introduced "electronic atoms of signal"\cite{Roussel:2016-2}, which
form a family of normed and mutually orthogonal electronic
wavelets representing the electron and hole wavefunctions generated
for each time period of the source. Generically, several electron/hole wavelets are needed, since several electron/hole excitations might be emitted in different wavepackets for each period of the source.
However, in our experimental situation at $f=\SI{9}{\giga\hertz}$, and more generally
whenever the probability to emit more than one electron/hole pair per period
is very small, only one electron $\varphi^{(e)}$ and one hole $\varphi^{(h)}$ wavelets are needed. The time translated
wavepackets $\varphi^{(\alpha)}_{l}(t)=\varphi^{(\alpha)}(t-lT)$
 are defined using the period index $l\in\mathbb{Z}$ and satisfy the following orthogonality conditions
$\langle\varphi^{(\alpha)}_{l}|\varphi^{(\beta)}_{l'}\rangle =
\delta_{\alpha,\beta}\delta_{l,l'}$ for $\alpha$ and $\beta$
being $e$ or $h$. The simplest generic form of the excess single electron coherence
can then be written in this case as:

\begin{widetext}
\label{eq:final-decomposition}
\begin{align}
\Delta_0\mathcal{G}^{(e)}_S(t,t')
=\sum_{(l,l')\in\mathbb{Z}^2}\left[g^{(e)}(l-l')\,
\varphi^{(e)}_{l}(t)\varphi^{(e)}_{l'}(t')^*
-g^{(h)}(l-l')\,\varphi^{(h)}_{l}(t)\varphi^{(h)}_{l'}(t')^* + 2\Re{\left(
g^{(eh)}(l-l')\,\varphi^{(e)}_{l}(t)\varphi^{(h)}_{l'}(t')^*
\right)} \right]
\end{align}
\end{widetext}

When $l=l'$, the real numbers
$0\leq g^{(e)}(0)\leq 1$ and $0\leq g^{(h)}(0)\leq 1$ represent the
probability for emitting an electronic (resp. hole) excitation with
wavefunction $\varphi^{(e)}$ (resp. $\varphi^{(h)}$) at each period.
When $l\neq
l'$, the complex numbers $g^{(e)}(l-l')$ (resp. $g^{(h)}(l-l')$)  represent the interperiod coherence between electronic (resp. hole)
wavepackets. Finally, the complex number $g^{(eh)}(l-l')$
represents the electron/hole coherence between the electron and hole
wavepacket associated with periods $l$ and $l'$. These terms occur when
the probability to emit the electron and hole excitations differ from 1,
such that a coherent superposition between the equilibrium state and the
creation of an electron/hole pair is generated. They encode the
modulus and phase of this coherent superposition. By exact
diagonalization of the projection of the excess single electron
coherence on the electronic and hole sectors (see Supplementary Material) it is
possible to extract from $W^{(e)}_S$ the
electron and hole wavefunctions  $\varphi^{(e/h)}$ as well as the
interperiod coherences $g^{(e)}(l-l')$, $g^{(h)}(l-l')$ and
$g^{(eh)}(l-l')$.  These data describe the single particle content of
the electronic current and quantify how far it deviates from the ideal
electron and hole emission regime, characterized in this language by
$g^{(eh)}(l-l')=0$ and $g^{(e)}(l-l')=g^{(h)}(l-l')=\delta_{l,l'}$.

\begin{figure}[h!]
\includegraphics[width=1 \columnwidth,keepaspectratio]{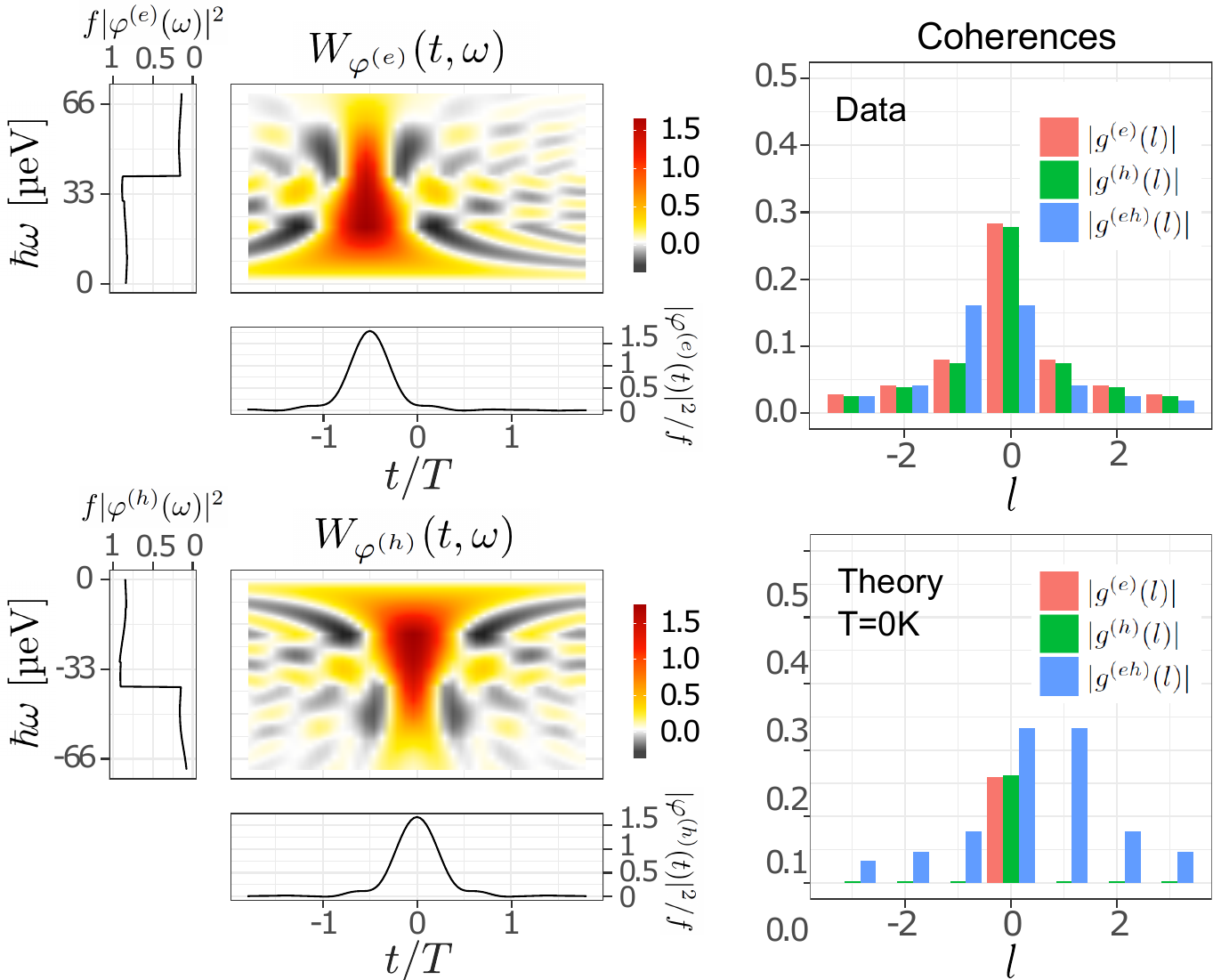}
\caption{\label{fig5}
 Left panel: Wigner distribution functions $W_{\varphi^{( e/h)}}(t,\omega)=\int d\tau \varphi^{(e/h)}(t+\frac{\tau}{2})\varphi^{(e/h) *}(t-\frac{\tau}{2}) e^{i\omega \tau}$ for the
dominant electronic  $\varphi^{(e)}$ and hole $\varphi^{(h)}$ atoms of signal in the $f=9~\mathrm{GHz}$ case. The panels in the margins of the colour plots represent the time $|\varphi^{(e/h)}(t)|^2/f$ and energy $f|\varphi^{(e/h)}(\omega)|^2$ distributions obtained by integration of $W_{\varphi^{( e/h)}}(t,\omega)$ over $\omega$ and $t$.  Right panel: moduli of the interperiod
coherence $|g^{(e)}(l)|$, $|g^{(h)}(l)|$ and $|g^{(eh)}(l)|$ (Data and numerical simulation at $T_{\text{el}}=\SI{0}{\milli\kelvin}$)}
\end{figure}

Figure \ref{fig5} presents the result of this
analysis on the experimental data obtained for the quantum drive ($f=\SI{9}{\giga\hertz}$). As expected, the
excess coherence is strongly dominated by one electronic $\varphi^{(e)}$ and one hole $\varphi^{(h)}$
wavepacket, which are plotted in the Wigner representation on the
upper and middle panels. The hole is shifted by half a period with respect to the
electron and its energy dependence is almost the same as that of the electron's at positive energy, as can be seen from
their electronic distribution functions
$f|\varphi^{(e/h)}(\omega)|^2$. Note that these functions present almost
flat plateaus of width $hf$: deviations from flatness express that atoms of signal at finite
temperature are contaminated by thermal excitations.
The moduli of the corresponding
interperiod coherences are depicted on the right panel: they extend
beyond one period, as expected, since the thermal coherence time  $h/k_BT_{\text{el}}\simeq \SI{0.5}{\nano\second}$ is roughly $4$ times
larger than the period. Since $g^{(eh)}(l-l')\neq 0$ and both
$g^{(e)}(0)\simeq g^{(h)}(0)\simeq 0.27\neq 1$, we are not in the single electron
regime. The specific role of thermal fluctuations is illustrated by comparing on Figure \ref{fig5} the data ($T_{\text{el}}=\SI{100}{\milli\kelvin}$) with the numerical simulation of the interperiod coherences in the zero temperature case. The occupation probabilities $g^{(e)}(0)$ and $g^{(h)}(0)$ increase from $0.16$ ($T_{\text{el}}=\SI{0}{\milli\kelvin}$) to $0.27$ ($T_{\text{el}}=\SI{100}{\milli\kelvin}$) when the temperature increases, reflecting the contamination by thermal excitations populating $\varphi^{(e/h)}$.
Thermal fluctuations also decrease the electron/hole coherence $|g^{(eh)}(0)|$, showing how temperature progressively destroys the coherent superposition between the equilibrium state and the creation of an electron/hole pair.

To conclude, we have demonstrated a quantum electrical current analyzer which directly extracts the single electron and hole wavefunctions, as well as their emission probabilities and coherence from one
emission period to the other. Assuming a
minimal knowledge on the state of the electron fluid, it can
be used to characterize any quantum electrical current.
It also explicitly takes into account the role of thermal excitations and their progressive contamination
of the electron and hole wavefunctions.
The same principle could lead to quantum current analyzers for
any low-dimensional conductor that can be weakly tunnel-coupled
to a probe port for noise measurements, enabling the control of the quantum
state of the elementary excitations transferred across nanoscale conductors.
Our quantum analyzer is the tool of choice for single electron source characterization or
for identifying single particle wavefunctions generated in
interacting conductors\cite{Marguerite2016}.
It may
also offer a way to access to the recently studied
electron/hole entanglement \cite{Hofer:2016-2} and,
supplemented by other measurements\cite{Thibierge2016}, to quantify
more precisely the importance of interaction-induced
quantum correlations. Finally, it can establish a bridge between electron and
microwave quantum optics \cite{Grimsmo:2016-1,Virally:2016-1}, by probing the electronic content of
microwave photons injected from a transmission line into
a quantum conductor.

This work has been supported by ANR grants
``1shot reloaded'' (ANR-14-CE32-0017), and ERC consolidator grant
``EQuO'' (No. 648236).

\clearpage
\onecolumngrid
\appendix


\section{Sample and noise measurements}

The sample is a GaAs/AlGaAs two dimensional electron gas of nominal
density $n_s=\SI{1.9e15}{\per\square\meter}$ and mobility
$\mu=\SI{2.4e6}{\per\square\centi\meter\per\volt\per\second}$.
It is placed in a magnetic field
$B=\SI{3.7}{\tesla}$ to reach the quantum Hall regime at filling factor $\nu \approx 2$ in the bulk.
The current noise at the output of the quantum point contact is converted to a voltage noise on the resistance $R=h/2e^2$ (at filling factor $\nu=2$)  between the ohmic contact connected to output 3 of the splitter and an ohmic contact connected to the ground. The contact on output $3$ is connected to
a tank circuit so as to move the noise measurement frequency to
$\SI{1.45}{\mega\hertz}$ with a $\SI{78}{\kilo\hertz}$
bandwidth. The tank circuit is followed by a pair of homemade cryogenic amplifiers
followed by room temperature amplifiers and a vector signal analyzer in order to
measure the voltage correlations between the two amplification chains. Noise measurements
are calibrated by measuring both the thermal noise of the output resistance $R$, and
the partition noise of a d.c. bias applied on input 2 of the electronic beam-splitter.

\section{Reconstruction of $\Delta W^{(e)}_{S,n}(\omega)$ from noise measurements}

The excess electronic distribution function $\Delta W^{(e)}_{S,n=0}(\omega)$ can be obtained via the derivative of $\Delta S$ with respect to the d.c. bias $\omega_{\text{DC}}$ applied on the probe port.
\begin{align}
\Delta S &= 2 e^2 \mathcal{T}(1-\mathcal{T}) \int
 \frac{\md\omega}{2\pi}\left[ \overline{\Delta W^{(e)}_S}^t
 \left(1-2f_{\text{eq}}\right)
 \right.   -\left.
 2\overline{\Delta W^{(e)}_S\Delta W^{(e)}_{P_n}}^t \right] \label{SHOM}\\
\widetilde{\Delta W}^{(e)}_{S,0} &= -\frac{\pi } {2 e^2
\mathcal{T}(1-\mathcal{T})} \frac{\partial \Delta S}{\partial
\omega_{\text{DC}}} = \int d\omega \Delta W^{(e)}_{S,0}\left(\omega\right)
\left(\frac{-\partial f_{\text{eq}}}{\partial \omega}\right)\left(\omega -
\omega_{\text{DC}}\right)
\label{n0}
\end{align}

As shown by Eq.\eqref{n0}, the experimental signal $\widetilde{\Delta W}^{(e)}_{S,0}$ does not provide
directly $\Delta W^{(e)}_{S,0}$ but its convolution with the thermally
broadened function $\left(\frac{-\partial f_{\text{eq}}}{\partial
\omega}\right)$. Knowing the electronic temperature and using deconvolution techniques based on Wiener filtering
\cite{Wiener} (see next section), one can reconstruct $\Delta
W^{(e)}_{S,0}(\omega)$ from the measurement of $\widetilde{\Delta
W}^{(e)}_{S,0}$.

The higher order terms $\Delta W^{(e)}_{S,n \neq 0}$ which encode all the time dependence, are reconstructed from the measurement of the output noise $\Delta S_{\phi}$ as a function of the phase $\phi$ of a.c. voltage $V_{P_n}(t)$ and of the d.c. voltage $\omega_{\text{DC}}$ applied on on the probe port. Using Eqs.\eqref{SHOM} with $\Delta
W^{(e)}_{P_n}(t,\omega)= -\frac{eV_{P_n}}{\hbar} \cos{\left(2 \pi n f t +
\phi\right)}  g_n\left(\omega -\omega_{\text{DC}}\right)$, we can reconstruct the real and imaginary parts of $\Delta W^{(e)}_{S,n}$:
\begin{subequations}
\begin{align}
 \Re{(\widetilde{\Delta W}^{(e)}_{S,n})} &=
 \frac{h}{8e^3V_{P_n}\mathcal{T}(1-\mathcal{T})} \left( \Delta
 S_{\phi=0}-\Delta S_{\phi=\pi} \right) = \int \md\omega \,
 \Re{\left(\Delta W^{(e)}_{S,n}\left(\omega\right)\right)} \,
 g_n(\omega-\omega_{\text{DC}}) \\
 \Im{(\widetilde{\Delta W}^{(e)}_{S,n})}
 &= \frac{h}{8e^3V_{P_n}\mathcal{T}(1-\mathcal{T})}   \left(\Delta
 S_{\phi=\frac{\pi}{2}}-\Delta S_{\phi=\frac{3\pi}{2}}\right)  = \int
 \md\omega \, \Im{\left(\Delta W^{(e)}_{S,n}\left(\omega\right) \right) }\,
 g_n(\omega-\omega_{\text{DC}})
\end{align}
\end{subequations}

As in the $n=0$ case, the experimental signal is the
convolution between $\Delta W^{(e)}_{S,n}$ and  $g_n=\Big(f_{\text{eq}}(\omega-n\pi f) - f_{\text{eq}}(\omega+n\pi f)\Big)/(2\pi nf)$. The real and imaginary parts of
$\Delta W^{(e)}_{S,n}$ are thus reconstructed using the Wiener filtering deconvolution
technique (see next section). A specific difficulty arises for the $n\neq 0$ terms, as their reconstruction process requires the
accurate knowledge of the relative phases between the probe signals for various values of $n$. To measure the phase relationship between the $n=1$, $n=2$ and $n=3$ probe signals, we use two-particle interferences between two different probe signals generated simultaneously at input $2$ in the absence of the source (e.g. between $n=1$ and $n=2$ or between $n=1$ and $n=3$). The noise at the splitter output is then minimal when the two probe signals are in phase, allowing for an accurate calibration of the relative phase between the probe signals for different harmonics $n$. For the measurement of the phase between the $n=2$ and $n=1$ harmonics, two-particle interferences between source and probe vanish at zero bias voltage $V_{\text{DC}}$ due to the electron/hole symmetry of a sinusoidal drive. We thus add a small positive bias voltage to the probe port in order to calibrate the relative phase (this method should be applied for the phase calibration of all even harmonics). As a result of the phase calibration, we find that, as theoretically expected for sine drives, $\widetilde{\Im{(\Delta W^{(e)}_{S,n})}}=0$ for all $n$ and all $\omega$. The reconstruction of $\Delta W^{(e)}_{S,n \neq 0}$ also requires the accurate knowledge of the probe voltage amplitude $V_{P_n}$ which are calibrated by measuring the partition
noise of the probe (source switched off) as a function of the drive amplitude $V_{P_n}$. When measuring these $n \neq 0$ harmonics, we also  systematically checked the linear dependence of the output noise with the probe amplitude in order to check the validity of the linear approximation relating $\Delta
W^{(e)}_{P_n}(t,\omega)$ to $V_{P_n}(t)$.

\section{Wiener deconvolution filter}

In order to account for the convolution of $\Delta W^{(e)}_{S,n}$
with $g_n$, the noise data $\widetilde{\Delta W}^{(e)}_{S,n}$ is deconvoluted
using a Wiener deconvolution filter \cite{Wiener}. Given our experimental situation, where $g_n$ is well known, but the experimental data $\widetilde{\Delta W}^{(e)}_{S,n}$ is spoiled by noise, brute force deconvolution techniques cannot be applied as they tend to amplify the noise fluctuations. On the contrary, the Wiener filter allows to efficiently deconvoluate the signal while smoothing the random fluctuations. Denoting by $FT$ and $FT^{-1}$ the Fourier and inverse Fourier transformations, which relate
the variable $\omega$ to its conjugate variable $\tilde{\omega}$, we define
the signal over noise ratio of the measurement:
$\rho(\tilde{\omega})  = \frac{\left|FT\left[g_n\right]\cdot FT\left[\Delta W^{(e)}_{S,n}\right]
\right|^2(\tilde{\omega})}{N(\tilde{\omega})}$, where $N(\tilde{\omega})$ is the  power spectral density
of the experimental noise. The Wiener filter that deconvolutes the
noisy experimental data is then defined as:
\begin{eqnarray}
\label{obviousdeconvolution}
\Delta W^{(e)}_{S,n}(\omega) & = &  FT^{-1}\left[H^*\left(\tilde{\omega}\right)\cdot
FT\left[\widetilde{\Delta W}^{(e)}_{S,n}\right](\tilde{\omega})\right] \\
H(\tilde{\omega}) & =&
\frac{1}{FT\left[g_n\right](\tilde{\omega})}\frac{1}{1+\rho^{-1}(\tilde{\omega})}
\end{eqnarray}
For $\tilde{\omega}$ values such that the signal to noise ratio is high
($\rho^{-1}(\tilde{\omega}) \ll 1$), the Wiener filter acts as a standard
deconvolution filter: $H=1/FT\left[g_n\right]$. On the contrary,
$\tilde{\omega}$ values which do not carry information on the signal
($\rho^{-1}(\tilde{\omega}) \gg 1$) are suppressed by the Wiener filter: $H \approx 0$,
so that contrary to standard deconvolution methods, the noise does not get
amplified by the deconvolution but is instead suppressed in the process.
The difficulty is then to properly evaluate $\rho(\tilde{\omega}) $.
Assuming no correlations between the fluctuations of $\Delta W^{(e)}_{S,n}$ measured at different energies $\omega$ (set by the d.c. bias $\omega_{\text{DC}}$), meaning that $N(\tilde{\omega})=N$ is a white noise, the value of $N$ is directly set by the size of the experimental error bars on the measurement of $\Delta S$. If the noise $N(\tilde{\omega})$ is known, it is not the case for the signal $FT\left[\Delta W^{(e)}_{S,n}\right](\tilde{\omega})$ which is the quantity to be determined. We have tested two different methods for
estimating $\rho$, which give equivalent results within error bars.
The first method relies on an estimate of the signal over noise ratio using a theoretical prediction
for $FT\left[\Delta W^{(e)}_{S,n}\right](\tilde{\omega})$. In the second method $\rho(\tilde{\omega}) $ is self-consistently determined from an optimization of the deconvolution process. More precisely, we take $\rho(\tilde{\omega})= \left|FT\left[g_n\right]\cdot FT\left[\widetilde{\Delta W}^{(e)}_{S,n}\right]
\right|^2(\tilde{\omega})/ a $, where $a$ is self-consistently determined. The optimization is based
on the balance between two limits. When $\rho $ is overestimated (meaning that the chosen $a$ is too small), the deconvolution
amplifies $\tilde{\omega}$ values which are in fact dominated by the noise which results in the appearance of high frequency fluctuations in the deconvoluted signal. When $\rho $ is underestimated (meaning that the chosen $a$ is too large),  $\tilde{\omega}$ values which are dominated by the signal are suppressed by the Wiener filter and the resulting deconvoluted data is oversmoothed. It can be detected by reapplying the convolution by
$g_n$ on the deconvoluted signal; oversmoothing then shows up as a
result which differs from the experimentally measured data $\widetilde{\Delta W}^{(e)}_{S,n}$.
This method leads to an uncertainty in the determination of $\rho $ (related to the uncertainty on the parameter $a$) which,
together with the noise on the experimental data, are taken into account when evaluating the
error bar of the deconvoluted data.
In order to evaluate these error bars, we generate a random distribution of experimental
data (with a standard deviation given by the experimental uncertainty) and
a family of Wiener deconvolution filters with randomly distributed values of $\rho$
(with a standard deviation given by the uncertainty on the determination of $\rho $).
We deconvoluate the randomly picked data with a randomly chosen filter,
generating a random distribution of deconvoluted data. The data points of $\Delta W^{(e)}_{S,n}$ plotted on
Fig.(2) and Fig.(3) correspond to the average of the obtained distribution
and the error bars to its standard deviation.

\section{Electron and hole wavefunction extraction}

The extraction of electron and hole wavefunctions from the experimental data of $\Delta W^{(e)}_{S}(t,\omega)$ relies on an algorithm which recasts any excess $T$-periodic single electron coherence
$\Delta_0\mathcal{G}^{(e)}$ under the form given by Eq.(4) of the article:
\begin{widetext}
\label{eq:final-decomposition}
\begin{align}
\Delta_0\mathcal{G}^{(e)}_S(t,t')
&=\sum_{(l,l')\in\mathbb{Z}^2}\left(g^{(e)}(l-l')
\,\varphi^{(e)}_{l}(t)\varphi^{(e)}_{l'}(t')^*
-g^{(h)}(l-l')\,\varphi^{(h)}_{l}(t)\varphi^{(h)}_{l'}(t')^*\right) \nonumber \\
&+\sum_{(l,l')\in\mathbb{Z}^2}\left(
g^{(eh)}(l-l')\,\varphi^{(e)}_{l}(t)\varphi^{(h)}_{l'}(t')^*
+\mathrm{h.c.}\right)
\end{align}
\end{widetext}

The algorithm is based on an exact diagonalization of the projection of $\Delta_0\mathcal{G}^{(e)}_S$ onto the electronic
and hole quadrants containing the information on the purely electronic
and purely hole single particle excitations within the system. Due to time periodicity,
this leads to electronic and hole probability spectra consisting of bands
on the quasi-pulsation interval $0\leq \nu\leq 2\pi f$ associated
with the time period $T$ as well as eigenvectors which are the Floquet
version of Bloch waves of solid state physics\cite{Roussel:2017-1}. The analogous of Wannier
functions\cite{Wannier:1937-1} are then the
electronic and hole atoms of signal\cite{Roussel:2016-2}. As in solid
state physics\cite{Marzari:2012-1}, the ambiguity of these wavepackets
has been lifted by minimizing their time spreading. Finally, the
electron and hole coherences $g^{(e)/h}(l)$ ($l\in\mathbb{Z}$) can be
obtained from the probability spectras and the electron/hole coherences
computed from $\Delta_0\mathcal{G}^{(e)}$ using the explicit numerical
data for the electronic and hole atoms of signals\cite{Roussel:2017-1}.

After Wiener filtering, the experimental data come as
a finite set of real values $\Delta
W^{(e)}_{S,n}(\omega_k)$ where $\omega_k$ are the discretized
values of $\omega$ and $n=0,\pm 1, \pm 2$. First we add the thermal
excess $f_{eq}(\omega)-\Theta(-\omega)$
of the equilibrium Fermi Dirac distribution at temperature
$T_{\text{el}}=100$~mK to $\Delta
W^{(e)}_{S,n=0}(\omega)$ to obtain the experimental dataset for $\Delta_0
W^{(e)}_{S}(\omega)$. Then,
in order to extract a square
matrix for the exact diagonalization, the first step is to generate new data that
can be matched on a grid well suited to the electronic and hole
quadrants. These two quadrants are defined in the frequency domain
as corresponding to the sectors where purely electronic (resp. purely
hole) excitation contribute to $\Delta
W^{(e)}_{S,n}(\omega)$. For a
periodically driven source, these sectors correspond to
$\omega\pm n\pi f\geq 0$ for the electron quadrant and
to $\omega \pm n\pi f\leq 0$ for the hole quadrant \cite{DegioWigner2013}. For each $n$, the dataset
$\Delta_0W^{(e)}_{S,n}(\omega_k)$
is first interpolated using cubic
splines to infer a new dataset on a grid adapted to the electronic and
hole quadrants, that is such that this grid intersects the boundaries
$\omega\pm n\pi f=0$ of the electronic and hole quadrants.
This new data grid has a discretization
step $\delta\omega$ such that $\hbar\delta\omega\simeq
\SI{0.19}{\micro\eV}$.
We assign zero to the
remaining grid points for which no data is available (higher harmonics
$|n|>2$ and values of $\hbar|\omega|\gtrsim \SI{70}{\micro\eV}$).
This dataset is then used to build the matrices
corresponding to the projection of this interpolated data for
$\Delta W^{(e)}_{S,n}(\omega)$, which are then fed into the electron and hole
wavefunction algorithm.

\end{document}